\documentstyle[preprint,aps]{revtex}
\tightenlines

\newcommand{\be}{\begin{equation}}
\newcommand{\ee}{\end{equation}}
\newcommand{\bea}{\begin{eqnarray}}
\newcommand{\eea}{\end{eqnarray}}

\def\scri{\hbox{${\cal J}$\kern -.645em 
{\raise.57ex\hbox{$\scriptscriptstyle(\ $}}}}

\begin{document}
\title{Gauge Independence and Relativistic Electron Dispersion Equation in Dense
Media }
\author{ D. Oliva Ag\"uero, H. P\'erez Rojas, A. P\'erez Mart\'\i nez and 
A. Am\'ezaga Hechavarr\'\i a}
\address{Centro de Matem\'aticas y F\'\i sica Te\'orica, Calle E No. 309, Ciudad\\
Habana, Cuba}
\maketitle

\begin{abstract}
We discuss the gauge parameter dependence of particle spectra in statistical
quantum electrodynamics and conclude that the electron spectrum is
gauge-parameter dependent. The physical spectrum being obtained in the
Landau gauge, which leads to gauge invariance in a restricted class of 
gauge transformations.
We compute the thermal self-energy of electrons in a dense
media to first order in the fine structure constant. In the
zero-temperature limit and at high density  the dispersion
equation is solved for some gauges, the curves differing from that
corresponding to the physical transverse gauge.
\end{abstract}

\section{Introduction}

The investigation of relativistic quantum electrodynamics at finite
temperature ($T$)\ and chemical potential ($\mu $) is of relevance in
several astrophysical and cosmological contexts\cite{Altherr}. Interest in
finite temperature and density of quantum field theories has increased
continuosly since the early seventies \cite{Linde}.
The problem about the dependence of the fermionic Green functions on the
gauge parameter is not new. For QED, it was solved by Landau and Khalatnikov
in \cite{Landau1} and simultaneously by Fradkin in \cite{Fradkin1}. These
authors showed that the longitudinal part of gauge field appears as a phase
factor into the electron Green function (see also \cite{Zumino}). The last
result was extended by Fradkin \cite{Fradkin2} to the statistical quantum
electrodynamics case, where the propagators also depend on the four-velocity
of the medium. 
It is a well established fact that in QFT the electron propagator is gauge
invariant on the tree-level mass shell ($P^2=m^2$), see \cite{weinberg}. For
finite temperature, this invariance has been also formally shown \cite
{landsman}. In the literature it has been explicitly verified this
conclusion in several limiting cases \cite{Toimela,Weldon1,Elmfors}.
However, it should be noted that the calculation of the mass operator is
carried out precisely to investigate the system perturbatively out of the
mass shell , where the condition of gauge parameter independence frequently
mentioned (see i.e. \cite{weinberg}) is not fulfilled, since the dispersion
equation differs in general from the form $P^2=m^2$, although it is expected
that the gauge independence of the dispersion equation remains valid near
the mass shell. But the departure of the spectrum from the $P^2=m^2$ form in
the temperature case \cite{Kapusta} excludes the validity of the above
mentioned proof at any order of perturbation theory. We will return to this
point along the paper.

A common practice in the literature has been to admit gauge independence, to
choose the Feynman gauge and to check it perturbatively in some limits.
Despite of the simplifications introduced by the Feynman gauge, the Landau
gauge has a special importance because the longitudinal part of the photon
propagator contributes as a phase factor to the electron Green function (see 
\cite{Fradkin2} for a general discussion) and thus it does not contribute to
the physical spectrum when the gauge parameter is chosen in a such a way
that the longitudial part of the photon propagator is zero.
The purpose of the present paper is to discuss the problem of
the gauge-parameter dependence in a non-perturbative finite temperature
field theory, 
to investigate the behavior of the electron dispersion equation at finite 
temperature and density and to consider the limit of $T=0$ and high density
afterwards.

This paper is organized as follows: in section $2$ we discuss the problem of
the gauge-parameter dependence in a non-perturbative finite temperature
field theory. In section $3$, we compute the electron self-energy in the one
loop approximation. In section $4$ two dispersion equations are obtained in
the limit of zero temperature for ultrahigh density. Section $5$ contain the
discussion and conclusions. In appendix A we derive the scalar quantities in
terms of which the mass operator is written, we give the definitions of the
main integrals that appear in the calculation, and we carry out the
Matsubara sums in the fourth momentum component $k_4$.

\section{Gauge-parameter dependence}

In what follows we want to argue in a perturbative-independent way that the
physical electron spectrum can be obtained (in a covariant way) only by using
the so-called Landau gauge $ 1/ \alpha=0$, which is actually compatible with 
a large class of gauge transformations, as indicated below. Usually the gauge
parameter dependence $\alpha$
is introduced in QED to make it possible to invert the (otherwise singular)
inverse photon propagator $D_{\mu \nu }^{-1}$ which is four-dimensional
transverse. But in QFT as well as in statistics, the photon spectrum is $\alpha
$-independent. In the case of a medium ($\mu \neq 0,\;T\neq 0$ ) due to the
gauge invariance of the theory \cite{Fradkin2} we can write the inverse
photon Green function, by introducing in the QED Lagrangian the gauge $\alpha
\partial _\mu A_\mu \partial _\nu A_\nu $ , as 
\begin{equation}
D_{\mu \nu }^{-1}=T_{\mu \nu }^1k^2-\Pi _{\mu \nu }+\alpha\,k_\mu k_\nu =0
\label{h1}
\end{equation}
where $T_{\mu \nu }^1=\left( \delta _{\mu \nu }-\frac{k_\mu k_\nu }{k^2}%
\right) $ and $\Pi _{\mu \nu }$ has the general four-dimensional tensor
structure 
\begin{equation}
\Pi _{\mu \nu }=T_{\mu \nu }^1A+T_{\mu \nu }^2B
\end{equation}
where 
\[
T_{\mu \nu }^2=\left( \frac{k_\mu k_\nu }{k^2}-\frac{k_\mu u_\nu }{(ku)}-%
\frac{u_\mu k_\nu }{(uk)}+\frac{u_\mu u_\nu k^2}{(ku)^2}\right) . 
\]
The photon Green function is then 
\begin{equation}
D_{\mu \nu }=T_{\mu \nu }^1A_1+T_{\mu \nu }^2B_1+
\frac{k_\mu k_\nu }{\alpha k^4}
\end{equation}
where
\begin{equation}
A_1=\frac 1{k^2-A},\,
B_1=\frac {B}{(k^2-A)( k^2-A+B(1-\frac{k^2}{(ku)^2}))}. 
\label{Dmedium}
\end{equation}
In the rest frame the polarization operator $\Pi _{\mu \nu }$ has the
tensor structure: 
\begin{equation}
\Pi _{\mu \nu }=\left\{ 
\begin{array}{l}
\left( \delta _{ij}-\frac{k_ik_j}{{\bf k}^2}\right) A\left( {\bf k}%
,k_4\right) +\Pi _{44}\!\frac{k_ik_j}{{\bf k}^2}\!\frac{k_4^2}{{\bf k}^2}%
\qquad \quad i,j=1,2,3 \\ 
\\ 
\Pi _{4i}=\Pi _{i4}=-\frac{k_ik_4}{{\bf k}^2}\Pi _{44}
\end{array}
\right.
\end{equation}
\noindent where $\Pi _{44}=\frac{{\bf k}^2}{k^2}\left( {}\right. A+\frac{%
{\bf k}^2}{k^2}B\left. {}\right) $. We have also  in the rest frame, in
the case $\alpha\to \infty $, 
\begin{equation}
D_{\mu \nu }=\left\{ 
\begin{array}{l}
\left( \delta _{ij}-\frac{k_ik_j}{{\bf k}^2}\right) C+\frac{k_ik_j}{{\bf k}^2%
}\frac{k_4^2}{{\bf k}^2}D_{44}\qquad \quad i,j=1,2,3 \\ 
\\ 
D_{4i}=D_{i4}=-\frac{k_ik_4}{{\bf k}^2}D_{44},
\end{array}
\right.  \label{Drest}
\end{equation}
\noindent 
where $C=1/(k^2-A)$, $D_{44}={\bf k}^4/[k^4({\bf k}^2-\Pi _{44})]$.

The photon spectrum may be obtained either by solving (\ref{h1}), in which $%
\alpha\neq 0$ factorizes, or from the poles of the four-dimensional transverse
part of $D_{\mu \nu }$. In the first case the following non-perturbative
gauge-parameter independent and gauge invariant spectrum is obtained in the
rest frame 
\begin{equation}
k^2+A({\bf k}^2,k_4)=0,\qquad 1+\frac{\Pi _{44}({\bf k^2},k_4)}{{\bf k}^2}=0
\label{h3}
\end{equation}
\noindent  The first equation corresponds to the spatial transverse modes
and the second (after multiplication by $k^2\neq 0$), to the spatial
longitudinal one. These are the physical modes, (which in the one-loop
approximation and at high temperature were studied in detail by \cite
{Weldon1}). A factor $k^4,$accounting for the unphysical modes introduced by
the gauge condition, when calculating $\left[ {}\right. DetD_{\mu \nu
}\left. {}\right] ^{-1/2}$, is removed by the Faddeev-Popov ghost term.
The photon spectrum can be also obtained directly from the poles of the 
Green function (\ref{Dmedium}). We have also the non-physical modes $k^4 =0$,
given by the longitudinal term, which are not present in (\ref{Drest}) since
they were removed by choosing the gauge parameter $\alpha\to \infty $.
In calculating 
the {\it exact} polarization operator in quantum field
theory as well as in statistics, only the transverse electron Green function
contributes.
The property stems from the gauge invariance and gauge parameter independence of
$\Pi_{\mu \nu}$, as  can be derived from the Ward identities
obtained from the
QED effective action $\Gamma$ in vacuum as well as in the temperature case, 
\begin{equation}
\partial_{\mu} (x_1)\frac{\delta^2 \Gamma(0)}{\delta A_{\mu} (x_1)\delta 
A_{nu} (x_2)} = \alpha\partial^2_{\nu} \partial_{\mu} \delta(x_1 - x_2)
\end{equation}
\noindent
which in momentum space reads
\begin{equation}
k_{\mu} D^{-1}_{\mu \nu} (k) = \alpha k_{\nu} k^2, \label{tras}
\end{equation}
\noindent where the substitution of (\ref{h1}) in (\ref{tras}) leads to the four-dimensional
transversality for $\Pi_{\mu \nu}$. This means that $\Pi _{\mu \nu }(x,y)$ is
gauge-invariant and gauge-parameter independent \cite{Fradkin2},\cite{Nash}.
Explicitly we have that $\Pi _{\mu \nu }(x,y)$ is given by
\begin{equation}
\Pi _{\mu \nu }(x,y)=e^2Tr\int \gamma _\mu G(x,z)\Gamma _\nu (z,y,y^{\prime
})G(y^{\prime },x)d^4xd^4y,  \label{Pol}
\end{equation}
\noindent
where the integration in the fourth coordinate
must be understood as in the interval $-\beta ,\beta $ in the temperature 
case, and $\Gamma(z,y,y^{\prime })=\gamma \delta (z-y)\delta (y-y^{\prime })-\frac{\delta
\Sigma (z,y)}{\delta ieA_\mu }$. 
Thus, if
a gauge parameter dependent $G(x,y|\alpha)$ is chosen, cancellations must
occur in the integrand in such a way that the integral obtained from (\ref
{Pol}) must be the same as that obtained by using $G(x,y\infty)=G(x,y).$ This
means that only the physical poles of $G(x,y)$ contribute to $\Pi _{\mu \nu }
$, and in consequence, to the transverse part of the exact photon Green
function $D_{\mu \nu }$.

Concerning the Fermion spectrum, the gauge parameter dependence is
introduced in the calculation of the mass operator $\Sigma$ in terms of $%
G(x,y |\alpha)$ and $D_{\mu \nu}(x,y |\alpha)$. Now, for the statistical case, as
shown by Fradkin \cite{Fradkin2}, the gauge parameter dependence is given
by a similar formula than in QED. We shall write it for the one-particle
electron Green function as,

\begin{equation}
G(x-y |\alpha)=G_0 (x-y)\exp \left\{ \,\,e^2\left[ \Delta ^l\left(
x-y\right) -\Delta ^l\left( 0\right) \right] \right\}  \label{k5a}
\end{equation}

\noindent where

\begin{equation}
\Delta ^l(x)=-\frac 1{(2\pi )^3\beta \alpha}\sum\limits_{k_4}\int \frac{d^3k}{%
k^4}{\large e}^{ikx}  \label{s2-11}
\end{equation}

\noindent
(in (\ref{k5a}), $G_0$ is the propagator in the Landau gauge ($1/ \alpha=0$) and
in our representation coincides with $G^t$), the poles of $G (p,\alpha)$ are
obtained from the poles of the Fourier transform of the right hand term in
(\ref{k5a}) which is the convolution of $G_0 (x-y)$ with the exponential $%
\alpha$-dependent factor. Due to this convolution, in any order of a
perturbative expansion of $G (p)$, the resulting poles are $\alpha$-dependent.

In addition to our previous arguments concerning the electron Green function
dependence on the gauge parameter, we want to mention the following one:
Let us call $\Sigma^t$ and $\Sigma^l$ respectively  the transverse and
longitudinal terms of the electron mass operator, where
\begin{equation}
\Sigma^l =\frac{e^2}{2\pi \beta}
\sum_{k_4}\int \gamma_{\mu} G(p+k)\Gamma_{\nu} (p+k, k)\frac{k_{\mu}k_{\nu}}{ k^2}D^l d^3 k,
\end{equation}
which is obviously nonzero and gauge-parameter dependent. Then, by taking
 $D^l (k) = 1/ \alpha k^2$ we observe it contains also
the contribution of the unphysical photon
modes $k^2_4 + {\bf k}^2 = 0$.
By using Ward 
identities it can be shown \cite{Fradkin2} 
that 
\begin{equation}
[i\gamma_{\mu}p_{\mu} + m + \Sigma^t(p)]G(p) = 1 - 
\frac{e^2}{2\pi \beta}
\sum_{k_4}\int \gamma_{\mu} G(p+k)\frac{k_{\mu}}{\alpha k^4}d^3 k. 
\label{sb}
\end{equation}
We observe also from this expression that the poles of $G(p)$ are given by the zeros of the term
in squared brackets in (\ref{sb}) plus the poles introduced by the second term containing the unphysical 
photon modes. 

Thus, from (\ref{s2-11}) or (\ref{sb}) it is seen that   the poles of the 
electron Green function become independent 
of the unphysical photon modes 
if it is taken the limit $\alpha$ $\to \infty $
words, by eliminating the contribution of these modes to
$\Sigma $. The zero
temperature QED
case can be treated by following similar arguments. We may cite especially ref.\cite{Lavelle}
in which a
very interesting discussion of the electron physical modes is given in the one-loop 
approximation for $\Sigma$ by using the Coulomb gauge.

Thus, although there are general proofs of gauge invariance for gauge boson
spectrum \cite{Kobes}, for fermions there are not (we consider there cannot
be) explicit proofs of gauge parameter independence of the one-particle
spectrum given by the poles of the Green function, except in very specific 
cases, as the infrared limit considered by
Abrikosov \cite{Abrikosov}. However, at least in the abelian case, if all
calculations are made in the Landau gauge $\alpha\rightarrow \infty $, which
corresponds to choosing the longitudinal part of the photon Green function
as equal to zero, the results lead to the physical fermion spectrum, and are 
{\it gauge invariant}  under the class of gauge transformations $A_\mu
(x)^{\prime }=A_\mu (x)+\partial _\mu \eta (x)$, $\psi (x)^{\prime
}=e^{e\eta (x)}\psi (x)$ in which  $\eta (x)=$ $- \partial _\mu A(x)_\mu /$ $%
\partial _\mu ^2$. This condition excludes the unphysical photon modes from
$\Sigma (x,y)$.

For non-abelian theories, an expression analogous to (\ref{k5a}) has not
been obtained yet. However, we consider the previous arguments might be
useful in considering the problem of gauge parameter independence of 
the fermion spectrum in the non-abelian case.

\section{Electron self energy}

In this section we will perform the computation of the mass operator in the
one loop approximation for a general gauge. We shall redefine  the gauge
parameter as $ \xi =1/ \alpha $ for simplicity  in the forthcoming expressions.
We will keep the dependence on
the gauge parameter to allow a further comparison of the results for
differents gauges. We start from the expression for the electron self-energy
in the one loop approximation in the temperature formalism (we use the
shifted momenta $p_\mu ^{*}=p_\mu -i\mu _e\delta _{4,\lambda }$)\cite
{Neutri1} 
\begin{equation}
\Sigma ^e(p)=\frac{e^2}{(2\pi )^3\beta }\sum_{k_4}\int d^3k\,\gamma _\mu
G^0(p^{*}+k)\Gamma _\nu ^0D_{\mu \nu }^0(k)  \label{k1}
\end{equation}
\begin{equation}
=\frac{e^2}{\left( 2\pi \right) ^3\beta }\sum\limits_{k_4}\int d^3k\left[
\gamma _\mu \frac 1{i(\gamma _\mu \,p_\mu ^{*}+k)^2+m_e^2}\gamma _\nu \frac{%
\delta _{\mu \nu }+\left( \xi -1\right) \,k_\mu k_\nu /k^2}{k^2}\right]
\end{equation}

Due to the medium we have the additional four velocity vector $u_\mu $ (in
the rest frame $u_\mu =(0,0,0,i)$, so that $k\cdot u=ik_4=-\omega $ . Here, $%
\xi $ is a parameter of arbitrary gauge) and thus the vectors, on which the
physical quantities may depend, are $k_\mu $ and $u_\mu $ The matrix
structure of the mass operator is now

\begin{equation}
\;\Sigma \left( p\right) =i\;(a\gamma _\rho p_\rho +b\gamma _\rho u_\rho +c)
\label{k5}
\end{equation}

After some work (see the appendix), we have, for an arbitrary $\xi $ 
\begin{equation}
a=g^2\frac 1{p^2}\left\{ \left( p^2I_1+4\,I_3+2\,I_4+2{\/}\,i\omega
\,I_5\,\right) +\xi \,\left( p^2I_1-\,I_3-2\,I_4-2{\/}\,i\omega
\,I_5\,\right) \right\}  \label{k6}
\end{equation}
\begin{eqnarray}
b &=&g^2\,\frac 1{p^2}\left\{ \left[ -3{\/}\,i\,\,p^2I_2-3\,\omega \,I_3+2{\/%
}\,\omega \,I_4-2{\/}\,i\left( p^2-\omega ^2\right) I_5+2\,\,p^2\omega
\,I_6\right] \right.  \nonumber \\
&&\ \left. +\,\xi \,\left[ -{\/}\,i\,\,p^2I_2+\,\omega \,I_3+2{\/}\,\omega
\,I_4-2{\/}\,i\left( p^2-\omega ^2\right) I_5+2\,\,p^2\omega \,I_6\right]
\right\}  \label{k7}
\end{eqnarray}
\begin{equation}
c=\left( 3+\xi \right) \,i\,g^2\,m_e\,I_1  \label{k8}
\end{equation}
where $g^2=e^2/\left( 2\pi \right) ^2$ is the fine structure constant and $%
I_1 - I_6$ are integrals defined in the appendix.

In the appendix, we evaluate these integrals, that is we perform the
Matsubara sum and do some algebra, subject to the following approximations:
(1) we drop the terms without dependence on the temperature, i.e. we ignore
non-thermal electrons; (2) we assume that the chemical potential is much
larger than the temperature, $\mu>>T$, so that we may approximate the
distribution function of the electrons by a step function \footnote{%
Actually, as soon as $\mu$ is about $10T$, this approximation, which is just
the characterization of the Fermi surface, is excellent, just as in
condensed matter.}; (3) in accordance with the previous point, since the
Fermi surface for positrons is negative, we assume that the distribution
functions for positrons are exactly zero; and (4) in agreement again with
all of this, we effectively neglect the temperature so that also the photon
distribution function is set to zero.

The upshot is a system with a Fermi liquid of electrons and no real photons
or positrons. Of course, in a physical situation charge must be balanced,
which in the framework we consider can be easily achieved through a proton
background (also, the proton and electron tadpoles cancel each other, so we
ignore tadpoles).

From the expressions for the integrals $I_i$ in the appendix, we carry out
the analytic prolongation $p_4+i\,\mu =i\omega $, the angular integrations,
and finally neglect the mass $\sqrt{\mu^2 -m_e^2} \simeq\mu$ in the upper
limit in the integrals. Note that we do not neglect the mass with respect to
the momentum. The expressions we obtain read as follows:

\begin{equation}
I_1=-\frac \pi p\int\limits_0^\mu \frac{dk\,k}{\sqrt{k^2+m_e^2}}\ln \left| 
\frac{A+B}{A-B}\right|  \label{k9}
\end{equation}
\begin{equation}
I_2=-\frac{\pi \,i}p\int\limits_0^\mu \frac{dk\,k\,\left( \omega -\sqrt{%
k^2+m_e^2}\right) }{\sqrt{k^2+m_e^2}}\ln \left| \frac{A+B}{A-B}\right|
\label{k10}
\end{equation}
\begin{equation}
I_3=-2\,\pi \int\limits_0^\mu \frac{dk\,k^2}{\sqrt{k^2+m_e^2}}\left[ -2+%
\frac AB\ln \left| \frac{A+B}{A-B}\right| -\frac{p^2}{2\,B}\ln \left| \frac{%
A+B}{A-B}\right| \right]  \label{k11}
\end{equation}
\begin{equation}
I_4=-2\,\pi \int\limits_0^\mu \frac{dk\,k^2}{\sqrt{k^2+m_e^2}}\left[ \frac{%
2\,p^2\left( k^2+p^2+A\right) }{A^2-B^2}-\frac{p^2}{2\,B}\ln \left| \frac{A+B%
}{A-B}\right| \right]  \label{k12}
\end{equation}
\begin{equation}
I_5=-2\,\pi \,i\int\limits_0^\mu \frac{dk\,k^2\,\left( \omega -\sqrt{%
k^2+m_e^2}\right) }{\sqrt{k^2+m_e^2}}\left[ \frac{A-2\,p^2}{A^2-B^2}-\frac 1{%
2\,B}\ln \left| \frac{A+B}{A-B}\right| \right]  \label{k13}
\end{equation}
\begin{equation}
I_6=-2\,\pi \int\limits_0^\mu \frac{dk\,k^2\,\left( \omega -\sqrt{k^2+m_e^2}%
\right) ^2}{\sqrt{k^2+m_e^2}\left( A^2-B^2\right) }  \label{k14}
\end{equation}
with the notation 
\begin{equation}
\left\{ 
\begin{array}{l}
A=p^2-\omega ^2-m_e^2+2\omega \sqrt{k^2+m_e^2} \\ 
B=2\,k\,p
\end{array}
\right.  \label{k15}
\end{equation}

\section{Dispersion Relations}

In a medium, the propagation of an electron is described by the effective
Lagrangian term 
\begin{equation}
L_m=\overline{\Psi }_e(p)\;S_e^{-1}\;\Psi _e(p) = \overline{\Psi }_e(p)\;
\left( i\,\gamma _\mu p_\mu ^{*}+m_e-{\bf \Sigma }\right) \; \Psi _e(p)
\label{k21}
\end{equation}

Accordingly, the equation of motion for the electron field is

\begin{equation}
S_e^{-1}\;\Psi _e(p)=0  \label{k22}
\end{equation}

The dispersion equation is thus 
\begin{equation}
\;\left| \gamma _\mu \left( \left( 1-a\right) \,p_\mu -i\,b\,\delta _{\mu
\,4}\right) -i\,\left( m_e-i\,c\right) \right| ^2=0  \label{k23}
\end{equation}
By the analytic continuation $ik_4=-\omega $, we get the equation for the
propagation of the electron in the medium as 
\begin{equation}
\sqrt{\left( 1-a\right) ^2p^2-\left( m_e-i\,c\right) ^2}-\left( 1-a\right)
\,\omega +b=0  \label{k24}
\end{equation}

In order to obtain analytic representations for the integrals $I_i$, we
neglect the electron mass in the square roots: this is justified because the
integrands are monotonously increasing in $k$ and the upper limit of
integration ($\simeq\mu $) is large. At ultrahigh densities, the electron
behaves like a massless fermionic quasiparticle. Let us illustrate the point
by sketching the evaluation of $I_1$ (\ref{k9}):

\begin{equation}
I_1=-\frac \pi p\int\limits_0^\mu dk\,\ln \left| \frac{p^2-\omega
^2-m_e^2+2\,k\,\left( \omega +p\right) }{p^2-\omega ^2-m_e^2+2\,k\,\left(
\omega -p\right) }\right|  \label{I1}
\end{equation}

Introduce the short-hand

\begin{eqnarray*}
\Delta &\equiv &p^2-\omega ^2-m_e^2
\end{eqnarray*}

\noindent whereby

\begin{equation}
I_1=-\frac \pi p\int\limits_0^\mu dk\,\left[ \ln \left| \Delta +2\,k\,\left(
\omega +p\right) \right| -\ln \left| \Delta +2\,k\,\left( \omega -p\right)
\right| \right]  \label{I2}
\end{equation}

Integrate by parts to obtain

\begin{eqnarray}
I_1 &=&-\frac \pi p\left[ \frac \Delta {2\,\left( \omega +p\right) }\ln
\left| \Delta +2\,\mu \,\left( \omega +p\right) \right| -\frac \Delta {%
2\,\left( \omega -p\right) }\ln \left| \Delta +2\,\mu \,\left( \omega
-p\right) \right| \right.  \nonumber \\
&&\quad \left. +\mu \ln \left| \frac{\Delta +2\,\mu \,\left( \omega
+p\right) }{\Delta +2\,\mu \,\left( \omega -p\right) }\right| +\frac{p\Delta 
}{\omega ^2-p^2}\ln \left| \Delta \right| \right]  \label{I3}
\end{eqnarray}

Now, take into account only the highest orders in $\mu $ (that is, assume $p$
small, $p<<\mu $) to obtain finally

\begin{equation}
I_1\approx -\pi \frac \mu p\ln \left| \frac{\omega +p}{\omega -p}\right|
+\cdot \cdot \cdot {,}  \label{k25}
\end{equation}

Similarly, we find 
\begin{equation}
I_2\approx \pi \,i\frac{\mu ^2}p\ln \left| \frac{\omega +p}{\omega -p}%
\right| +\cdot \cdot \cdot {,}  \label{k26}
\end{equation}
\begin{equation}
I_3\approx \pi \,p\,\mu \ln \left| \frac{\omega +p}{\omega -p}\right| +\cdot
\cdot \cdot {,}  \label{k27}
\end{equation}
\begin{equation}
I_4\approx \frac \pi 2\frac{p^2}{\omega ^2}\mu ^2+\cdot \cdot \cdot {,}
\label{k28}
\end{equation}
\begin{equation}
I_6\approx -\frac 14\frac{\mu ^2}{\omega ^2}+\cdot \cdot \cdot \text{.}
\label{k28b}
\end{equation}

The analysis for $I_5\approx 0$ is simpler because for $p\rightarrow 0$ we
have from (\ref{k23}) 
\begin{equation}
I_5\sim \left[ \frac{\Delta +2\,\omega \,k-2\,p^2}{\left( \Delta +2\,\omega
\,k\right) ^2-(2\,k\,p)^2}-\frac 1{4\,k\,p}\ln \left| \frac{\Delta
+2\,k\,\left( \omega +p\right) }{\Delta +2\,k\,\left( \omega -p\right) }%
\right| \right]
\end{equation}
\begin{equation}
\rightarrow \quad \left[ \frac 1{\Delta +2\,\omega \,k}-\frac 1{4\,k\,p}%
\left( \frac{4\,k\,p}{\Delta +2\,\omega \,k}\right) \right] =0  \label{k28c}
\end{equation}

we obtain, for the ultrahigh density limit ( $\mu >>m_e$ ), the following
expresions for the functions $a$, $b$ y $c$:

\begin{equation}
a=g^2\pi \left[ \left( 1-\xi \right) \frac{\mu ^2}{\omega ^2}-\left(
3+2\,\xi \right) \frac \mu p\ln \left| \frac{\omega +p}{\omega -p}\right|
\right] ,  \label{k29}
\end{equation}

\begin{equation}
b=g^2\pi \left[ \left( 3+\xi \right) \frac{\mu ^2}{2\,p}\ln \left| \frac{%
\omega +p}{\omega -p}\right| -\frac{\mu ^2}{2\omega }-\xi \frac{\omega \mu }p%
\ln \left| \frac{\omega +p}{\omega -p}\right| \right] ,  \label{k30}
\end{equation}

\begin{equation}
c=-\left( 3+\xi \right) \,g^2\pi \,i\,m_e\frac \mu p\ln \left| \frac{\omega
+p}{\omega -p}\right| \text{.}
\end{equation}

Replacing these expressions for the integrals in the dispersion equation (%
\ref{k24}), is easy to obtain the dispersion relations, which constitute our
main result: 
\begin{equation}
\omega _{+}=\sqrt{p^2+m_e^2}+g^2\,\pi \,\frac{\mu ^2}{2\,p}\left[ \left(
1-\xi \right) \frac{\,p}{\omega _{+}}+\left( 3+2\,\xi \right) \ln \left| 
\frac{\omega _{+}+p}{\omega _{+}-p}\right| \right]  \label{k32}
\end{equation}

\begin{figure}[p]
\centerline{\input{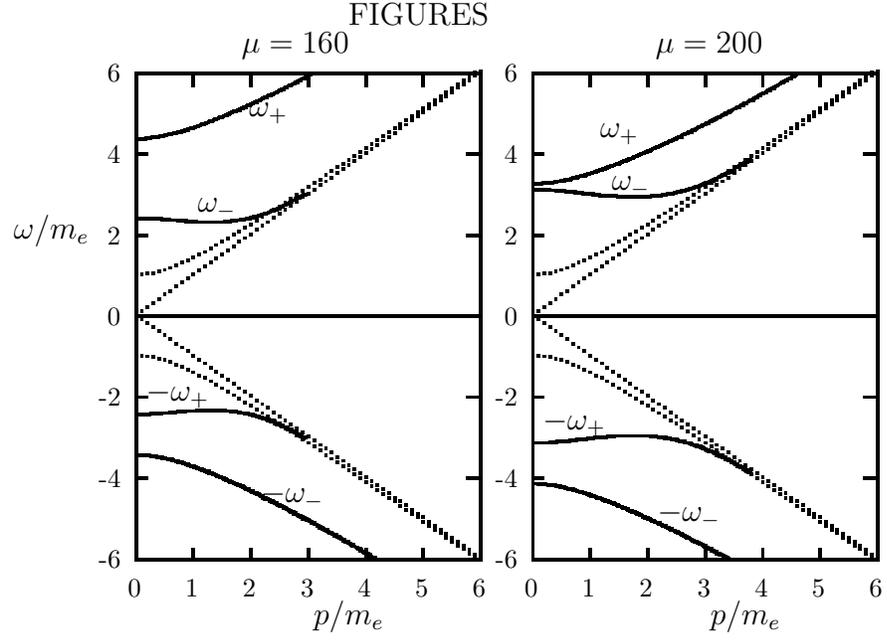}}
\caption{The spectrum of fermionic excitations in an 
ultrarelativistic plasma for two values of density. In dots 
are showed the light cone and the free dispersion relation.}
\label{xfig3}
\end{figure}

Here, $\omega _{+}$ is the normal dispersion relation and corresponds to an
electron-like quasi-particle. There is also a abnormal dispersion branch ($%
\omega _{-}<0$), which have an essential collective character:

\begin{equation}
\omega _{-}=-\sqrt{p^2+m_e^2}+g^2\,\pi \,\frac{\mu ^2}{2\,p}\left[ \left(
1-\xi \right) \frac{\,p}{\omega _{-}}+\left( 3+2\,\xi \right) \ln \left| 
\frac{\omega _{-}+p}{\omega _{-}-p}\right| \right]  \label{k33}
\end{equation}
The solution of (\ref{k32}) and (\ref{k33}) are showed in Fig.\ref{xfig3}
together with the light cone and the free dispersion law for two differents
values of density. As can be observed the behavior of the dispersion curves
is analogous to the reported in \cite{blaizot}, although in our case lower
densities are employed, which allows us to have a more realistic description
of the astrophysical and cosmological contexs. In Figs.\ref{xfig4}and \ref
{xfig5} are plotted the normal and abnormal dispersion curves for three
values of the density. In the case of $\omega _{+}$ it is seen that its
effective mass grows and its approach to the light cone is faster with
increasing density. In the case of $\omega _{-}\ $,its curve tends to
approach the light cone faster than $\omega _{+}$, moreover it can be
observed the occurence of a minimum for a finite value of $p$ which
decreases with the density. The abnormal branch shows a negative effective
mass near the $p=0$ as the hole quasiparticles.

Note that we get two different dispersion relations for particles and holes,
unlike other results at finite temperature, such as densities near the
electron mass \cite{Levinson}, high temperature but without chemical
potential \cite{Toimela}, and others \cite{Petitgirard}. By taking into
account the electron mass, we correct the pathological behavior of the
derivative of the dispersion curve at $p=0$ obtained by several authors for
massless fermions \cite{Weldon1,Weldon2,Altherr,Elmfors,Braaten}.

\section{Discussion}

We return to the problem of gauge invariance. As shown in Figs.\ref{xfig1}, 
\ref{xfig2}, the spectrum is dependent on the gauge parameter, but for $%
p/m_e > 1$ and near the mass shell, the dispersion curves obtained, for some
values of the gauge parameter, approach among themselves and to the mass
shell and become gauge-independent. This in some way verifies the mass shell 
$P^2=m^2$ gauge independence pointed out by several authors, but we observe
it occurs for large momenta, i.e. in the light cone region. But for some
other gauges the behavior is quite different. In (\ref{xfig2}) one can see
that in the case of the gauge $\xi =-3$ the dispersion curve behaves in a
drastic different way as the cases $\xi =1$ and $\xi =0$ in the same region
of momenta. One must stress at this point that the gauge-dependent scalar $b 
$ plays an essential role in producing a departure of the spectrum from the $%
P^2 = m^2$ mass shell, which manifests especially in the low momentum limit: 
$\lim_{p \to 0} b = b(\omega, \mu, \xi) \neq 0$. Thus, the mass shell gauge
invariance of $\sum$ \cite{Toimela} does not lead to a gauge-independent
spectrum. From the above results we conclude that the information extracted
from the curves in the region far from the mass shell, e.g. the effective
mass, are not physical but a gauge artifact. However the results of section
2 indicate that the Landau gauge, in despite of introducing algebraic
complications, is the appropriate.

The minimum for $p_0\neq 0$ in the abnormal solution, suggests an analogy
with a superfluid behavior, (or superconductivity, since we are considering
charged fermionics particles), but this conclusion can't obtained only of
mass operator. To give a criterium of superconductivity is necessary to
analyze the vertex part. The electrons at high density constitute a weakly
interacting attractive Fermi system. Obviously, to reach the conclusion that
this system exhibits superfluidity, it is better to shift the energy axis to
the point $m_e+\delta m(\mu )$, where $\delta m(\mu
)=\lim\limits_{p\rightarrow 0}\left( \omega -m_e\right) $. There, the
positron-like quasiparticle spectrum is tangent at $p=0$ (as in the
non-relativistic case \cite{Landau2}).

On the other hand, the tangent to the positron-like quasiparticle curve
gives the velocity below which a superfluid effect can be expected.

This astonishing behavior of the dispersion curve for the positron-like
quasiparticle, in a weakly coupled Fermi gas, was already observed in a very
different context by Weldon in 1989 \cite{Weldon2}, who considered a
quark-gluon plasma at very high temperature and without chemical potential.
There, superfluidity is not intuitive at all. We think that the analogy with
cold helium can be established with more sense in our case, of degenerate
fermion gas at low temperature, since the phenomena of superconductivity and
superfluidity disappear at high temperatures.

\section{Acknowledgments}

It is a pleasure to acknowledge many enlightening discussions with A. Cabo,
M. Ruiz-Altaba and M. Torres.

\section{Appendix}

To obtain the constants on which the mass operator depends, we proceed as in
reference \cite{Fradkin1}. First, define the auxiliar functions 
\begin{equation}
A=\frac 14Tr\left[ \sum \left( p\right) \cdot \gamma ^\mu p_\mu \right]
\label{A1}
\end{equation}
\begin{equation}
B=\frac 14Tr\left[ \sum \left( p\right) \cdot \gamma ^\mu u_\mu \right]
\label{A2}
\end{equation}
\begin{equation}
C=\frac 14Tr\left[ \sum \left( p\right) \right]  \label{A3}
\end{equation}
Using (\ref{k5}), we find 
\begin{equation}
a=A+i\,p_{4\,}^{*}B  \label{A4}
\end{equation}
\begin{equation}
b=\frac 1{p^2}\left( i\,p_4^{*}\,A-p^{*2}\,B\right)  \label{A5}
\end{equation}
\begin{equation}
c=C  \label{A6}
\end{equation}

The main integrals that appear in the calculation of these constants are 
\begin{equation}
I_1=\sum_{k_4}\int \!{\frac 1{\left( {k}^2+k_4^2\right) \left( {k}%
^2+k_4^2+2\,kp+2\,k_4\,p_4+{m}^2+{p}^2+p_4^2\right) }}{d}^3{k}  \label{A7}
\end{equation}
\begin{equation}
I_2=\sum_{\ k_4}\int \!\frac{k_4}{\left( k^2+k_4^2\right) \left(
k^2+k_4^2+2\,kp+2\,k_4\,p_4+m^2+p^2+p_4^2\right) }d^3k{\,\,}  \label{A8}
\end{equation}
\begin{equation}
I_3=\sum_{k_4}\int \!{\frac{k\,p}{\left( {k}^2+k_4^2\right) \left( {k}%
^2+k_4^2+2\,kp+2\,k_4\,p_4+{m}^2+{p}^2+p_4^2\right) }}{d}^3{k}  \label{A9}
\end{equation}
\begin{equation}
I_4=\sum_{k_4}\int \!{\frac{k\,p^2}{\left( {k}^2+k_4^2\right) ^2\left( {k}%
^2+k_4^2+2\,k\,p+2\,k_4\,p_4+{m}^2+{p}^2+p_4^2\right) }}{d}^3{k}  \label{A10}
\end{equation}
\begin{equation}
I_5=\sum_{k_4}\int \!{\frac{k\,\,p\,k_4}{\left( {k}^2+k_4^2\right) ^2\left( {%
k}^2+k_4^2+2\,k\,\,p+2\,k_4\,p_4+{m}^2+{p}^2+p_4^2\right) }}{d}^3{k}
\label{A11}
\end{equation}
\begin{equation}
I_6=\sum_{k_4}\int \!{\frac{k_4^2}{\left( {k}^2+k_4^2\right) ^2\left( {k}%
^2+k_4^2+2\,kp+2\,k_4\,p_4+{m}^2+{p}^2+p_4^2\right) }}{d}^3{k}  \label{A12}
\end{equation}
The integration is over $k$-three-space and the Matsubara sum runs over the
discrete fourth component of momentum, which for bosons is taken as $%
k_4=2n\pi \beta $ ($n\in {\bf Z}$).

Assuming $\mu>>T$, only thermal electrons survive with a step-function
distribution, whereas both positrons and photons get killed, and we find

\begin{equation}
I_1=\int \frac{{\bf \theta }\left( \mu -\varepsilon _{k+p}\right) }{%
2i\varepsilon _{k+p}\left[ -p_4-i\left( \varepsilon _k-\varepsilon
_{k+p}+\mu \right) \right] \left[ -p_4+i\left( \varepsilon _k+\varepsilon
_{k+p}-\mu \right) \right] }d^3k  \label{A13}
\end{equation}

\begin{equation}
I_2=\int \frac{\left[ -p_4+i\left( \varepsilon _{k+p}-\mu \right) \right] 
{\bf \theta }\left( \mu -\varepsilon _{k+p}\right) }{2i\varepsilon
_{k+p}\left[ -p_4-i\left( \varepsilon _k-\varepsilon _{k+p}+\mu \right)
\right] \left[ -p_4+i\left( \varepsilon _k+\varepsilon _{k+p}-\mu \right)
\right] }d^3k  \label{A14}
\end{equation}

\begin{equation}
I_3=\int \frac{\overrightarrow{k}\,\cdot \overrightarrow{p}{\bf \theta }%
\left( \mu -\varepsilon _{k+p}\right) }{2i\varepsilon _{k+p}\left[
-p_4-i\left( \varepsilon _k-\varepsilon _{k+p}+\mu \right) \right] \left[
-p_4+i\left( \varepsilon _k+\varepsilon _{k+p}-\mu \right) \right] }d^3k
\label{A15}
\end{equation}

\begin{equation}
I_4=-\int \frac{\left( \overrightarrow{k}\,\cdot \overrightarrow{p}\right) ^2%
{\bf \theta }\left( \mu -\varepsilon _{k+p}\right) }{2i\varepsilon
_{k+p}\left[ \left( p_4+i\mu \right) +i\left( \varepsilon _k-\varepsilon
_{k+p}\right) \right] ^2\left[ \left( p_4+i\mu \right) -i\left( \varepsilon
_k+\varepsilon _{k+p}\right) \right] ^2}d^3k  \label{A16}
\end{equation}

\begin{equation}
I_5=-\int \frac{\overrightarrow{k}\,\cdot \overrightarrow{p}\,\left[ -\left(
p_4+i\,\mu \right) +i\,\varepsilon _{k+p}\right] \,{\bf \theta }\left( \mu
-\varepsilon _{k+p}\right) }{2i\varepsilon _{k+p}\left[ \left( p_4+i\,\mu
\right) +i\left( \varepsilon _k-\varepsilon _{k+p}\right) \right] ^2\left[
\left( p_4+i\,\mu \right) -i\left( \varepsilon _k+\varepsilon _{k+p}\right)
\right] ^2}d^3k  \label{A17}
\end{equation}

\begin{equation}
I_6=-\int \frac{\,\left[ -\left( p_4+i\,\mu \right) +i\,\varepsilon
_{k+p}\right] ^2\,{\bf \theta }\left( \mu -\varepsilon _{k+p}\right) }{%
2i\varepsilon _{k+p}\left[ \left( p_4+i\,\mu \right) +i\left( \varepsilon
_k-\varepsilon _{k+p}\right) \right] ^2\left[ \left( p_4+i\,\mu \right)
-i\left( \varepsilon _k+\varepsilon _{k+p}\right) \right] ^2}d^3k
\label{A18}
\end{equation}

In the above, $\varepsilon _{k+p}=\sqrt{\left( k+p\right) ^2+m^2}$, $%
\varepsilon _k=\left| \overrightarrow{k}\right| $and ${\bf \theta }\left(
\lambda \right) $ is the Heaviside unitary step function characteristic of
the low temperature limit.

\newpage
\newpage
\begin{figure}[h]
\centerline{\input{figura2.tex}}
\caption{Dependence of the normal branch with density. 
The dots lines correspond to the light cone and the free 
dispersion relation.}
\label{xfig4}
\end{figure}
\vspace{2cm}
\begin{figure}[h]
\centerline{\input{figura3.tex}}
\caption{Dependence of the abnormal branch with density.
The dots lines correspond to the light cone and the free 
dispersion relation.}
\label{xfig5}
\end{figure}
\newpage
\begin{figure}[h]
\centerline{\input{figura1.tex}}
\caption{Dependence of the normal branch with the
 gauge parameter. The dots and dashed lines 
correspond to the light cone and the free 
dispersion relation respectively.}
\label{xfig1}
\end{figure}
\vspace{2cm}
\begin{figure}[h]
\centerline{\input{figura4.tex}}
\caption{Dependence of the abnormal branch with the
gauge parameter. The dots and dashed lines 
correspond to the light cone and the free 
dispersion relation respectively.}
\label{xfig2}
\end{figure}
\end{document}